\documentclass[twocolumn,showpacs]{revtex4}
\usepackage{graphicx,amssymb}
\begin{document}

\title{Theory of carrier mediated ferromagnetism in dilute magnetic oxides}

\author{M.J. Calder\'on, S. {Das Sarma}}  
\affiliation{Condensed Matter Theory Center, Department of Physics,
  University of Maryland, College Park, Maryland 20742-4111}

\begin{abstract}
We analyze the origin of ferromagnetism as a result of carrier mediation
in diluted magnetic oxide semiconductors in the light of the
experimental evidence reported in the literature. We propose that a
combination of percolation of magnetic polarons at lower temperature
and Ruderman-Kittel-Kasuya-Yosida ferromagnetism at higher temperature may be the reason for
the very high critical temperatures measured (up to~$\sim 700$K).
\end{abstract}

\maketitle

\section{Introduction}
Semiconductors doped with magnetic ions are being
studied in an effort to develop spintronics, the new kind of electronics that
seeks to
exploit, in addition to the charge degree of freedom as in the usual electronics, also the spin of the carriers~\cite{zutic-review}. The first so-called diluted magnetic semiconductors (DMS) were II-VI semiconductor alloys like Zn$_{1-x}$Mn$_x$Te and Cd$_{1-x}$Mn$_x$Te~\cite{furdyna88} originally studied in the 1980s. These materials are either spin glasses or have very low ferromagnetic (FM) critical temperatures $T_C$ ($\sim$ few K)~\cite{ferrand01} and are, therefore, inadequate for technological applications which would require FM order at room temperature. 
 More recently, the Mn doped III-V semiconductors In$_{1-x}$Mn$_x$As~\cite{munekata89,ohno92} and  Ga$_{1-x}$Mn$_x$As~\cite{ohno96,jungwirthRMP06} showed ferromagnetism at a much higher temperature, thanks to the development of molecular beam epitaxy (MBE)-growth techniques. The current high $T_C$ record of $173$K achieved in Mn-doped GaAs by using low temperature annealing techniques~\cite{wang02,edmonds02,chiba03} is promising, but still too low for actual applications. In all these materials, ferromagnetism has been proven to be carrier mediated, a necessary property for spintronics since this enables the modification of magnetic behavior through charge manipulation. 
This has motivated a search for alternative spintronics materials with even higher $T_{\rm C}$ and carrier mediated FM. 
In this direction, dilute magnetic oxides~\cite{spaldin-review}, 
such as magnetically-doped
TiO$_2$~\cite{matsumoto01}, ZnO~\cite{ueda01}, and
SnO$_2$~\cite{ogale03}, could represent this alternative with reported $T_{\rm C}$s above room temperature and as high as $\sim 700$K~\cite{shinde03}. 

The general formula for these oxide based dilute magnetic semiconductors (O-DMS) is
\begin{equation}
A_{1-x}M_x O_{n-\delta} \, , \nonumber
\end{equation}
where $A$ is a non-magnetic cation, $M$ is a magnetic cation, 
$\delta$ is the concentration of oxygen vacancies which depends on the
growth conditions, and $n= 1$ or
$2$. Carriers, usually electrons, are provided by the oxygen vacancies that are believed to act as shallow
donors~\cite{forro94,tang94}. This is in contrast to III-V semiconductors, like Ga$_{1-x}$Mn$_x$As, where the carriers (holes in this case) are provided by the magnetic impurities themselves which act also as donors (or acceptors).
More oxygen vacancies and, thus, more
carriers, are produced at low oxygen pressure~\cite{toyosaki04,venkatesan04}. 
In the process of doping with magnetic ions, usually of different
valence than the ion they substitute for, oxygen or $A$ vacancies are also
introduced to maintain charge neutrality. However, these oxygen vacancies have
been found not to contribute directly to the electrical conductivity of the
system~\cite{chambers03}.  On the contrary, the resistivity
increases by orders of magnitude upon doping~\cite{ogale03,shinde03}.

There is
currently no consensus on the origin of ferromagnetism in O-DMS, 
in particular, whether it is an extrinsic effect due to direct interaction between the local moments in magnetic impurity clusters (or nanoclusters) or is
indeed an intrinsic property caused by exchange coupling between
the spin of the carriers and the local magnetic moments. This is a very important issue because spintronics requires the carriers to be polarized and this would only be guaranteed if ferromagnetism is intrinsic. Experimental evidence for carrier-mediated FM in O-DMS is not yet conclusive. 
In the much
studied Co-doped TiO$_2$ in the rutile phase, anomalous Hall
effect (AHE)~\cite{toyosaki04} and electric field induced modulation
of magnetization (by as much as $13.5\%$)~\cite{zhao-FE-05}  have been observed, arguing for carrier-mediated FM. However, AHE was also measured in a sample with magnetic
clusters, casting doubts about the conclusions that can be drawn from
AHE data~\cite{shinde-HE-04}.  Nevertheless, room temperature FM has also been reported in cluster free films grown or annealed at high temperatures $\sim 900 ^{o}$C ~\cite{shinde03} by pulsed laser deposition, and in nanocrystalline films  grown in conditions that
preclude the formation of metallic cobalt~\cite{chambers04}.
Another piece of
evidence
in favor of carrier-mediated ferromagnetism is the observation that
the magnetic field dependence of ferromagnetic circular dichroism
is in good agreement with those of magnetization and anomalous Hall
effect~\cite{toyosaki05}. More recently, X-ray photoemission
spectroscopy measurements have suggested strong hybridization between
carriers in the Ti 3d band and the localized t$_{2g}$ states of Co$^{2+}$~\cite{quilty06}.

Many reports have raised serious doubts on the magnetism of magnetically-doped ZnO~\cite{rao05} since the results are very sensitive to sample
preparation. However, it has been pointed out~\cite{rao05,spaldin04} that
the lack of ferromagnetism in some samples can be the result of too low a density of
carriers. Indeed, films of Zn$_{0.75}$Co$_{0.25}$O prepared in low 
oxygen partial pressure ($< 10^{-6}$ Torr), a condition that should 
increase the density of
electrons from oxygen vacancies, were found to be ferromagnetic at 
room temperature~\cite{rode03}.  These samples exhibited perpendicular
magnetic anisotropy~\cite{rode03,dinia05} and no segretion effects
were significant. In the same direction, doping Zn$_{0.98}$Co$_{0.02}$O with small amounts of
Cu also enhanced the system's ferromagnetism~\cite{lin04}. The
systematic variation of magnetism in doped ZnO as a function of the
magnetic dopant has been explored in Ref.~\cite{venkatesan04},
where room-temperature ferromagnetism has been found in films doped
with Sc, Ti, V, Fe, Co, or Ni but not with Cr, Mn, or Cu. The same
group also reported an increase in  
the magnetic moment per Co ion as a function of the reduction of
oxygen pressure (equivalent to an increase of the density of
carriers)~\cite{venkatesan04}. Optical magnetic circular dichroism,
one of the tests proposed as a signature of diluted ferromagnetism~\cite{zhao-FE-05}, has
also been measured at low temperatures~\cite{ando01}, and at room temperature~\cite{neal06}.
Very recently, by analyzing the controlled introduction and removal of interstitial Zn, carrier-mediated ferromagnetism
in Co-doped ZnO has been demonstrated~\cite{kittilstved06}.

The case of magnetically-doped SnO$_2$ 
(with $T_C \sim 600$K~\cite{ogale03,coey04}) 
seems to be different from the
previous two materials in that the parent compound is highly conducting~\cite{chopra83}
(though still transparent)
and that some doped samples have shown an extremely large magnetic moment
($>7 \mu_B$)
per magnetic impurity~\cite{ogale03,coey04}. 

It thus appears that different doped magnetic oxides may very well have different underlying mechanisms leading to the observed ferromagnetism -- in particular, there may very well be several competing FM mechanisms in play in O-DMS, a novel theme we develop further here.

In this article, we address the origin of ferromagnetism as mediated by carriers in O-DMS
analyzing the existing experimental evidence for magnetic and
transport properties, in particular, in Co-doped TiO$_2$. Our goal is to develop a theory for O-DMS ferromagnetism (assuming it to be mediated by band or impurity-band carriers) in analogy with the better understood III-V Mn-doped DMS materials~\cite{jungwirthRMP06,dietlreview02,timm03,dassarmaSSC03,macdonaldNatMat}. 
Magnetically doped III-V semiconductors are ferromagnetic for a wide range
of carrier concentrations, from the insulating to highly conducting
regimes, where
different mechanisms are expected to prevail. It has been
proposed that for high enough values of the local exchange between the
carriers and the magnetic ions, an impurity band is
formed~\cite{chatto01,calderon02}.  In a highly
insulating system, the Fermi level is well below the mobility
edge of the impurity band. In this regime,
ferromagnetism has been explained as the result of percolation of bound magnetic
polarons~\cite{kaminski02}. This mechanism is
consistent with the concave shaped magnetization versus temperature
$M(T)$ curves observed, for instance, in (In,Mn)As~\cite{ohno92}. In
the more conducting samples, itinerant carriers would mediate
ferromagnetism via a Ruderman-Kittel-Kasuya-Yosida (RKKY) mechanism whose sign fluctuations, and the associated
frustration effects, are
suppressed due to the low density of carriers~\cite{priour04}, arising
possibly from heavy compensation and/or existence of defects.

Given that the carriers in magnetic oxides reside in an insulating impurity band, there are essentially three kinds of carrier-mediated magnetic exchange interactions which could potentially lead to intrinsic carrier-mediated ferromagnetism: double exchange (similar to the situation in manganites)~\cite{zener} in the impurity band, bound magnetic polarons percolation (similar to the situation in insulating diluted magnetic semiconductors, e.g. Ge$_{1-x}$Mn$_x$, In$_{1-x}$Mn$_x$As, and low-$x$ Ga$_{1-x}$Mn$_x$As) ~\cite{kaminski02}, and indirect (RKKY) exchange coupling mediated by free carriers (similar to the situation in the optimally doped $x \approx 0.05$ high-$T_{\rm C}$ metallic Ga$_{1-x}$Mn$_x$As)~\cite{priour04}.  The double exchange mechanism gives, at the low carrier density of magnetic oxides,  $T_{\rm C}$ proportional to the carrier density; therefore, critical temperatures exceeding room temperature ($\sim 300 K$) are essentially impossible within this model. 
In general, the carrier-mediated indirect RKKY exchange mechanism applies only to free carriers in an itinerant band, and therefore may be thought to be ruled out for doped magnetic oxides which are insulating impurity band materials. But we propose here that the 'standard' RKKY mechanism may very well be playing a role in the magnetic oxides, particularly at high temperatures, where a large number of localized impurity band carriers will be thermally activated (due to the small values of the carrier binding energies in the magnetic oxides) either within the impurity band or to the conduction band, effectively becoming itinerant free carriers (albeit thermally activated ones) which can readily participate in the indirect RKKY exchange between the localized impurity magnetic moments. If the exchange coupling between local moments and the thermally activated carriers is sufficiently large, then this mechanism could explain the very high $T_C$ observed in magnetic oxides. We emphasize that $T_C$ also depends on other parameters  including the activated carrier density and the magnetic moment density. At low temperature, the RKKY mechanism must freeze out since thermal activation from the impurity band is no longer operational and one must therefore have a complementary mechanism to provide carrier-mediated ferromagnetism. We believe that the bound magnetic polaron  (BMP) mechanism is the plausible low temperature magnetic ordering mechanism, but it cannot explain the high (room temperature or above) claimed $T_C$ of O-DMS unless one uses completely unphysical parameters.

Our new idea in this work is, therefore, the suggestion that the intrinsic carrier-mediated ferromagnetism leading to high Curie temperatures is plausible in doped magnetic oxides only if two {\em complementary} magnetic mechanisms (i.e. the bound magnetic polaron percolation at low temperatures and the indirect RKKY exchange mechanism in the presence of substantial thermal activation of carriers in the impurity band) are operating in parallel. We find that no single carrier-mediated mechanism, by itself, can account for the observed high $T_C$ in doped magnetic oxides. These considerations apply
mainly to TiO$_2$ as high-$T$ resistivity measurements indicate existence of substantial thermally activated carrier population~\cite{shinde03,toyosaki04,higgins-HE-04}. Of course, the possibility that the observed ferromagnetism in magnetic oxides arises from completely unknown extrinsic mechanisms (e.g. clustering near the surface) or from non-carrier-mediated mechanisms cannot be ruled out at this stage. Only more experimental work, perhaps motivated by our theoretical considerations presented herein, can provide definitive evidence in favor of one or more FM mechanisms in O-DMS. We mention in this context that even for much better-understood III-V DMS systems there are still debates in the literature regarding the precise role of the RKKY mechanism~\cite{priourPRL06}.

This paper is organized as follows: in Sec.~\ref{sec:model}, the bound magnetic polaron and the RKKY model are
introduced. In Sec.~\ref{sec:compare}, 
magnetic and transport properties
of dilute magnetic oxide semiconductors, mainly Co-doped TiO$_2$, 
are summarized and analyzed
in the light of our proposed combined model. Sec.~\ref{sec:discussion}
presents a discussion of our model and some of the alternatives suggested in
the literature. We conclude in Sec.~\ref{sec:conclusion}. 

\section{Model}
\label{sec:model}
The general Hamiltonian that describes the physics of diluted magnetic semiconductors is (see, for instance, Ref.~\cite{dassarma-mag-03})
\begin{eqnarray}
H&=& \sum_{\alpha} \int{ d^3 x \,\, \Psi_{\alpha}^{\dagger} (x) \left(-{\frac{\nabla^2}{2 m}} + V_L(x)+ V_r(x) \right) \Psi_{\alpha} (x)} \nonumber\\
&+&\int{ d^3 x \sum_{i\alpha}  W(x-R_i) \Psi_{\alpha}^{\dagger} (x) \Psi_{\alpha} (x) }\nonumber\\
&+& \int{d^3 x \sum_{i\alpha\beta}  J {\mathbf S}_i \,\, \Psi_{\alpha}^{\dagger} (x)  {\boldmath\sigma}_{\alpha,\beta} \Psi_{\beta} (x) \,\,a_0^3 \,\delta^3(x-R_i)  } \nonumber \\
&+&\sum_{i,j} J_d (R_i-R_j) {\mathbf S}_i {\mathbf S}_j 
\label{eq:hamiltonian}
\end{eqnarray}
where $m$ is the relevant effective mass, $V_L$ is the periodic lattice potential, $V_r$ is a potential arising from disorder (magnetic and non-magnetic) in the lattice, $W$ is a Coulombic potential arising from the oxygen vacancies that act as shallow donors, $J$ is the local exchange (Hund's like) between the carrier spin and the magnetic impurities moments, and $J_d$ is a direct exchange between the magnetic impurities spins (which is ferromagnetic in Co-doped TiO$_2$~\cite{janisch06}). ${\mathbf S}_i$ is the impurity spin located at ${\mathbf R}_i$, ${\boldmath \sigma}_{\alpha,\beta}$ represents the Pauli matrices with spin indices $\alpha$ and $\beta$, and $a_0^3$ is the unit cell volume. We are neglecting the electron-electron interaction which we expect to be very small due to low carrier density in the diluted magnetic oxides. In the following, we assume the direct exchange term to be effectively included in the local exchange term. Only at very low carrier densities and in the case of antiferromagnetic $J_d$, which is {\it not} the case of Co-doped TiO$_2$, the direct exchange could compete with the carrier-mediated FM and cause frustration~\cite{kaminski04}. 

The Hamiltonian in Eq.~(\ref{eq:hamiltonian}) is extremely complex to solve exactly and, consequently, we integrate out the electronic degrees of freedom and simplify the problem by considering only the term in the Hamiltonian that dictates the interaction between the spin of the carriers and the magnetic moments
\begin{equation} 
H_l=\sum_i J a_0^3 \,\,\,{\mathbf S}_i \,{\mathbf s}({\mathbf R}_i)\,.
\label{eq:hamiltonian-local}
\end{equation}
We, therefore, model the problem with a minimum set of parameters that effectively include other interactions in the system.
This simplified Hamiltonian is solved in two complementary cases: (i) localized carriers (bound to oxygen vacancies by the interaction $W$), and (ii) itinerant carriers. 
When the carriers are localized, they form bound magnetic
polarons that percolate at $T_C$~\cite{kaminski02,coey05}. 
When carriers are delocalized in the conduction (or valence) band, they can mediate ferromagnetism through the RKKY mechanism. In the following, we introduce these two approaches, which are capable of producing carrier-mediated FM provided the carrier density is much lower than the magnetic impurity density.

\subsection{Percolation of bound magnetic polarons}

\begin{figure}
\begin{center}
        \includegraphics*[width=3.0in]{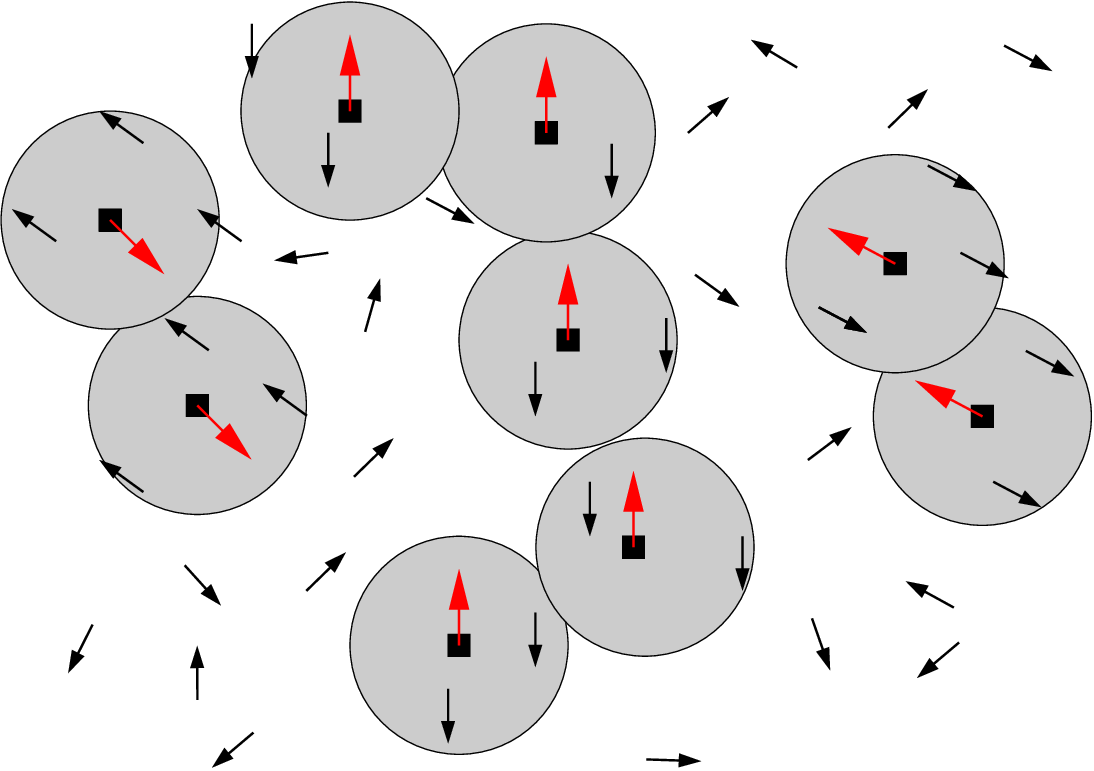}
        \end{center}
        \caption{Schematic representation of magnetic percolation in
        oxide based 
        dilute magnetic semiconductors. The
        solid squares represent the oxygen vacancies where an
        electron, represented by an arrow, is localized. The gray
        circles represent the extension of the electron
        wave-function. Magnetic
        impurity spins are represented by small arrows whose
        orientation is established by antiferromagnetic exchange coupling with the localized carrier. }
\label{fig:percolation}
\end{figure}

Bound magnetic polarons (BMP) are the result of the
combination of Coulomb and magnetic exchange interactions~\cite{kasuya68}. 
Carriers are localized due to electrostatic interaction
with some defect (i.e. the magnetic impurity in III-V semiconductors, and the oxygen vacacies in the magnetic oxides) with a confinement radius $a_{\rm B}=
\epsilon (m/m^*) a$, with $\epsilon$ the static dielectric constant, $m^*$ the
effective mass of the polaron, and $a=0.52 \,{\rm\AA}$  the Bohr radius. $a_{\rm B}$ is larger (smaller) for shallower (deeper) defect levels. We assume the bound electron (or hole) wave-function has the isotropic hydrogen-like form $\Psi(r) \sim {\frac{1}{\sqrt{a_{\rm B}^3}}}\exp(-r/a_{\rm B})$ so that the exchange [given in Eq.~(\ref{eq:hamiltonian-local})] between the magnetic impurities and the carrier decays exponentially with the distance $r$ between them as $\sim \exp(-2r/a_{\rm B})$. At a certain $T$, the radius of the polaron $R_{\rm p}$ is given by the condition 
\begin{equation}
k_{\rm B} T= |J| \left(a_0/a_{\rm B}\right )^3 S s \exp(-2R_{\rm p}/a_{\rm B}) \,,
\end{equation}
where $k_{\rm B}$ is the Boltzmann constant, leading to 
\begin{equation}
R_{\rm p}(T) \equiv (a_{\rm B}/2) \ln \left(sS|J|\left(a_0/a_{\rm B}\right )^3/k_{\rm B}T\right) \,,
\label{eq:pol-radius}
\end{equation}
where we can see that at the high temperature  disordered phase there are no polarons, which start forming at $k_B T \sim sS|J|\left(a_0/a_{\rm B}\right )^3$, and then grow as temperature is lowered. The impurity spins that are a distance $r<R_{\rm p}$ from a localized carrier tend to align with the localized carrier spin. The polarized magnetic impurities form, in turn,  a
trapping potential for the carrier such that a finite energy is required
to flip its spin. The magnetic polarons are well
defined non overlapping isolated entities only at low carrier densities and
sufficiently large temperatures. The size of
the polarons increases as temperature decreases, eventually overlapping with neighboring BMPs. 
 This overlap causes the alignment of their spins, therefore forming FM clusters. The FM transition takes place  when an 'infinite cluster' (of the size of the system) is formed, i.e. when the BMP percolation occurs.  
This scenario has been
studied in Ref.~\cite{kaminski02} in the context of III-V semiconductors. There it is shown that
the bound magnetic polaron model can be mapped onto the problem of percolation of overlapping spheres.  
The model proposed is valid
in the low carrier density
regime $n_c a_{\rm B}^3 <<1$ and when the density of magnetic impurities
$n_i$ is larger than the density of carriers $n_c$. Under these
conditions, each
carrier couples to a large number of magnetic impurities, as shown in Fig.~\ref{fig:percolation}. 

In order to calculate the BMP percolation critical temperature $T_C^{\rm perc}$ we first need to estimate
the maximum temperature at which two magnetic polarons a distance $r$ from each other are still strongly correlated $T_{2\rm p} (r)$. By estimating the number of impurities which interact with both polarons~\cite{kaminski02}, this temperature is found to be given by

\begin{equation}
k_{\rm B} T_{2p}(r) \sim a_B \sqrt{r n_i} \,s S J  \left({{a_0}\over{a_{\rm B}}} \right )^3 \exp(-r/a_{\rm B})\,.
\label{eq:t2p}
\end{equation}

As the temperature is lowered more and more polarons overlap with each other until a cluster of the size of the sample appears. The critical polaron radius at which this happens can be calculated as the percolation radius in the problem of randomly placed overlapping spheres. This problem has been solved numerically~\cite{pike74} giving
\begin{equation}
r_{\rm perc}\approx 0.86/\sqrt[3]{n_c} \,.
\end{equation}
Substituting this distance in Eq.~\ref{eq:t2p} gives the FM transition temperature
\begin{equation}
k_{\rm B} T_C^{\rm perc} \sim s S J \left({{a_0}\over{a_{\rm B}}} \right )^3 \left(a_{\rm B}^3 n_c
  \right) ^{1/3} \sqrt{{{n_i}\over{n_c}}} \,\exp \left(- {{0.86}\over {\left(a_{\rm B}^3 n_c
  \right) ^{1/3} }}\right).
\label{eq:Tc-perc}
\end{equation}

The magnetization is due to the magnetic ions (since $n_i > n_c$) in the percolating
cluster. It is given by 
\begin{equation}
M(T)= S n_i \,\mathcal{V} \left( r_{\rm corr} \sqrt[3]{n_c} \right)
\end{equation}
where $\mathcal{V} (y)$ is the infinite cluster's volume in the model of the overlapping spheres
and 
\begin{equation}
r_{\rm corr}(T)=\left[0.86+(a_B^3 n_c)^{1/3} \ln {\frac{T_C}{T}}\right]/\sqrt[3]{n_c}
\end{equation}
is the maximum distance between correlated polarons at $T$. Therefore, the shape of the magnetization curve only depends on $a_B^3 n_c$.
The volume of the infinite cluster is strongly suppressed as
temperature is increased producing a concave shaped $M(T)$ curve~\cite{kaminski02}, as
opposed to the Weiss mean-field like convex Brillouin function shape
expected for  
highly conducting samples. The
concave shape is less pronounced for larger $a_{\rm B}^3 n_c$~\cite{dassarma-mag-03}, that would eventually lead to a change of
shape in the itinerant carrier regime, beyond the limit of
applicability of this theory. Interestingly,
however, the BMP percolation theory developed in the strongly localized
insulating regime smoothly extrapolates to the mean-field RKKY-Zener theory
(discussed below in Sec.~\ref{sec:model}B) in the itinerant
free-carrier metallic regime~\cite{kaminski02,dassarma-mag-03}, giving us some additional confidence in the potential validity of the composite percolation-RKKY
model we are proposing here for ferromagnetism in the magnetic oxides
materials.

Conduction in a bound magnetic polaron system occurs via an activated
process. Thus, $\rho \sim \exp(\Delta E/k_{\rm B} T)$ where $\Delta E$ is the
activation (or binding) energy.  $\Delta E$  depends on the binding energy of
the carrier due mainly to electrostatic interaction with donors or
acceptors, and in a minor degree, on the polarization due to magnetic exchange
Eq.~(\ref{eq:hamiltonian-local}). The effect of an applied magnetic field 
in this system is to align non-connected polarons in such a way that
the polarization part of the binding energy gets suppressed.
As a result, the magnetoresistance
is negative and depends exponentially on the field~\cite{kaminski03}.
It should be emphasized that the BMP picture and the associated
magnetic percolation transition is only valid in the strongly
localized carrier regime where each BMP (i.e. each carrier) is
immobile, and a percolation transition in the random disordered
configuration makes sense.

\subsection{RKKY model for itinerant carriers}

As already mentioned, carriers in magnetic oxides are donated by oxygen vacancies that
act as shallow donors. 
The carrier binding energy 
has been estimated for different samples of undoped TiO$_2$ 
to be $\sim 4$ meV~\cite{forro94}, $\sim 14$ meV~\cite{tang94} and
$\sim 41$ meV~\cite{shinde-un}. These are smaller than the
Mn acceptor binding energy of GaAs, which is of the order
of $0.1$ eV~\cite{yakunin04}.
It is then possible that
a large fraction of the electrons are promoted (i.e. thermally activated) above the mobility
edge in the impurity band at temperatures lower than $T_C^{\rm perc}$ and that the
localization picture just described cannot explain ferromagnetism by
itself. This thermal activation of carriers will be extremely
effective at higher temperatures, particularly since $k_{\rm B} T_C$ in this
system could be substantially higher than the electron binding energy
or, equivalently, the activation energy. The assumption of strict BMP
localization fails in this situation as the carriers get activated
into highly mobile states at higher temperatures, as manifested by the
experimentally observed enhancement in the conductivity at higher
temperatures. The BMP formation will be suppressed in this situation since the thermally activated band carriers are effectively free or itinerant at elevated temperatures.

Itinerant carriers coupled to local moments by Eq.~(\ref{eq:hamiltonian-local})
lead to the well-known Zener-RKKY mechanism of indirect magnetic interaction between the magnetic
impurities. This model gives an effective exchange $J_{\rm eff}$ between magnetic moments in the lattice of the form~\cite{ruderman54,yosida57}
\begin{equation}
J_{\rm eff} \propto {{2 k_F R_{ij}  \cos(2 k_F R_{ij})- \sin(2 k_F R_{ij})}\over {R_{ij}^4}}
\label{eq:RKKY-osc}
\end{equation}
where $R_{ij}=R_i-R_j$ (i.e. the distance between magnetic impurities) and $k_F$ is the Fermi momentum. $J_{\rm eff}$ oscillates in sign with distance leading in general to complicated magnetic configurations~\cite{mattis62}, frustration and glassy behavior. However, in the limit of very low density of carriers relevant here ($n_c << n_i$), $k_F r \rightarrow 0$ and  $J_{\rm eff} <0$, namely, the RKKY interaction is always ferromagnetic.
The RKKY ferromagnetism phase diagram has recently been calculated~\cite{priourPRL06}.

We study the RKKY interaction in the limit of non-degenerate carriers within
mean-field theory (therefore, all disorder in the lattice is neglected). The interaction between carrier spins and local moments
can be then described as a self-consistent process in which carrier spins
see an effective magnetic field produced by the local moments (which are considered classical)
\begin{equation}
B_{\rm eff}^{(c)}=\frac{Ja_0^3n_i \langle S_z \rangle}{g_c \mu_B}\,,
\end{equation}
and the local
moments see an effective field produced by the carrier spins
\begin{equation}
B_{\rm eff}^{(i)}=\frac{Ja_0^3n_c \langle s_z \rangle}{g_i \mu_B}\,.
\end{equation}

The response of the impurity spin, which follows Boltzmann statistics, to $B_{\rm eff}^{(i)}$ is~\cite{ashcroft} 
\begin{equation}
\langle S_z \rangle = S\, \mathcal{B}_S \left({{g_i \mu_B B_{\rm eff}^{(i)}}\over{k_B T}}\right)=S\, \mathcal{B}_S \left({{S J a_o^3 n_c \langle s_z
    \rangle}\over{k_{\rm B} T}} \right)
\label{eq:Sz}
\end{equation}
where
\begin{equation}
\mathcal{B}_s(y) \equiv {\frac{2s+1}{2s}} \coth {\frac{2s+1}{2s}} y  -{\frac{1}{2s}} \coth {\frac{1}{2s}} y  
\end{equation}
is the Brillouin function. In the nondegenerate limit, the carrier spin distribution is not affected by the Pauli exclusion principle and, therefore, it is determined also by Boltzmann statistics rendering
\begin{equation}
\langle s_z \rangle =s \,\mathcal{B}_s \left({{g_c \mu_B B_{\rm eff}^{(c)}}\over{k_B T}}\right)= s \, \mathcal{B}_s \left({{s J a_o^3 n_i \langle S_z
    \rangle}\over{k_{\rm B} T}} \right)
\label{eq:sz}
\end{equation}
Combining Eqs.~(\ref{eq:Sz}) and ~(\ref{eq:sz}), we get a self-consistent equation for the impurity spin
\begin{equation}
\label{eq:RKKY}
\langle S_z \rangle  = S\,\mathcal{B}_S \left[{\frac{J a_0^3}{k_B T}} s  \mathcal{B}_s \left({\frac{J a_0^3 n_i \langle S_z \rangle}{k_B T}}\right)\right]\, .
\label{eq:RKKY-mag}
\end{equation}
When the effective magnetic fields are small, as it is close to the magnetic transition, we can perform the expansion of $\mathcal{B}_s(y)$ for small values of $y$
\begin{equation}
\mathcal{B}_s(y) \approx {\frac{s+1}{3s}}y+O(y^3) \, ,
\end{equation}
which, applied to Eq.~\ref{eq:RKKY}, gives the critical
temperature~\cite{coey-comment}

\begin{equation}
k_{\rm B} T_C^{RKKY} ={{1}\over{3}} \,J a_0^3 \,\sqrt{n_c n_i}\, \sqrt{(S+1) (s+1)}.
\label{eq:Tc-nondeg}
\end{equation}

The magnetization is mainly due to the local magnetic moments $\langle
S_z \rangle$ (because $n_i>n_c$) and is given by the self-consistent Eq.~\ref{eq:RKKY-mag}. The resultant dependence with
temperature is concave for low values of $n_c/n_i$
 and convex mean-field-like for $n_c/n_i
\ge 0.2$~\cite{dassarma-mag-03}.

For higher carrier density, the carriers form a degenerate gas 
and the procedure
followed in  Ref.~\cite{priour04} would be more appropriate.
In this reference, the RKKY model is solved 
taking into account the disorder in the lattice. The conductivity is included
via the mean free path (MFP), which produces a cutoff in the RKKY
coupling reach.  $T_C$ has been found
to depend very strongly on the relation between $n_c$, $n_i$ and the
MFP. MFP can be made larger by improving sample quality. 
In contrast with mean field treatments of the same model (which neglect disorder) that
predict $T_C \propto n_c^{1/3}$~\cite{dietl01},
$T_C$ is enhanced and later suppressed by increasing $n_c$. 
This $T_C$ suppression is due to the sign oscillations of the RKKY interaction  at larger carrier densities [see Eq.~(\ref{eq:RKKY-osc})] that lead to magnetic frustration and spin glassy behavior.
It is
then proposed~\cite{dassarma04} that $T_C$ improvement can arise both
from increasing $n_c$ and the MFP. 
The magnetization curves $M(T)$ are convex in the highly conducting
limit (large MFP) and
concave for more insulating systems (small MFP)~\cite{priour04,dassarma-mag-03} appropriate for dilute magnetic oxides.
This is qualitatively the same result as given by the non-degenerate
approach and the BMP model.
In this way, the
localized and itinerant carrier models predict the same
behavior for $M(T)$ in both the high and low carrier density limits,
leading to correct qualitative predictions even beyond the
applicability range of the two models.

Calculations of dc-resistivity in diluted systems 
within DMFT approximation have rendered
a negative magnetoresistance MR, that peaks at $T_C$ but is appreciably smaller than the MR
of magnetic ordered lattices such as manganites~\cite{hwang-condmat}.

\subsection{Double exchange model}
In the double exchange model~\cite{zener}, a large $J$ forces the spin of the carrier to be parallel to the local magnetic moment in such a way that the kinetic energy of the carriers hopping between magnetic sites is minimized when the magnetic ions are ferromagnetically ordered. 
This model was proposed for manganites, where ferromagnetism and metallicity usually come together. The ferromagnetic
critical temperature in the double exchange model is proportional to the density of
carriers and the bandwidth~\cite{calderon-MC98}, with very small
values in the low density limit appropriate to O-DMS. 
Therefore, double exchange is not a
suitable mechanism to explain the high critical temperature of dilute
magnetic oxides, and is hence not considered further in this work. However, in the strong coupling regime, where the effective exchange coupling $J$ is large ($J>>t, \, E_F$ where $t$ is the carrier bandwidth), a double exchange mechanism may be appropriate for low-$T_C$ DMS materials as discussed in Refs.~\cite{chatto01,akaiPRL98}. 
   
\section{Comparison to experiment}
\label{sec:compare}

\begin{figure}
\begin{center}
        \includegraphics*[width=3.0in]{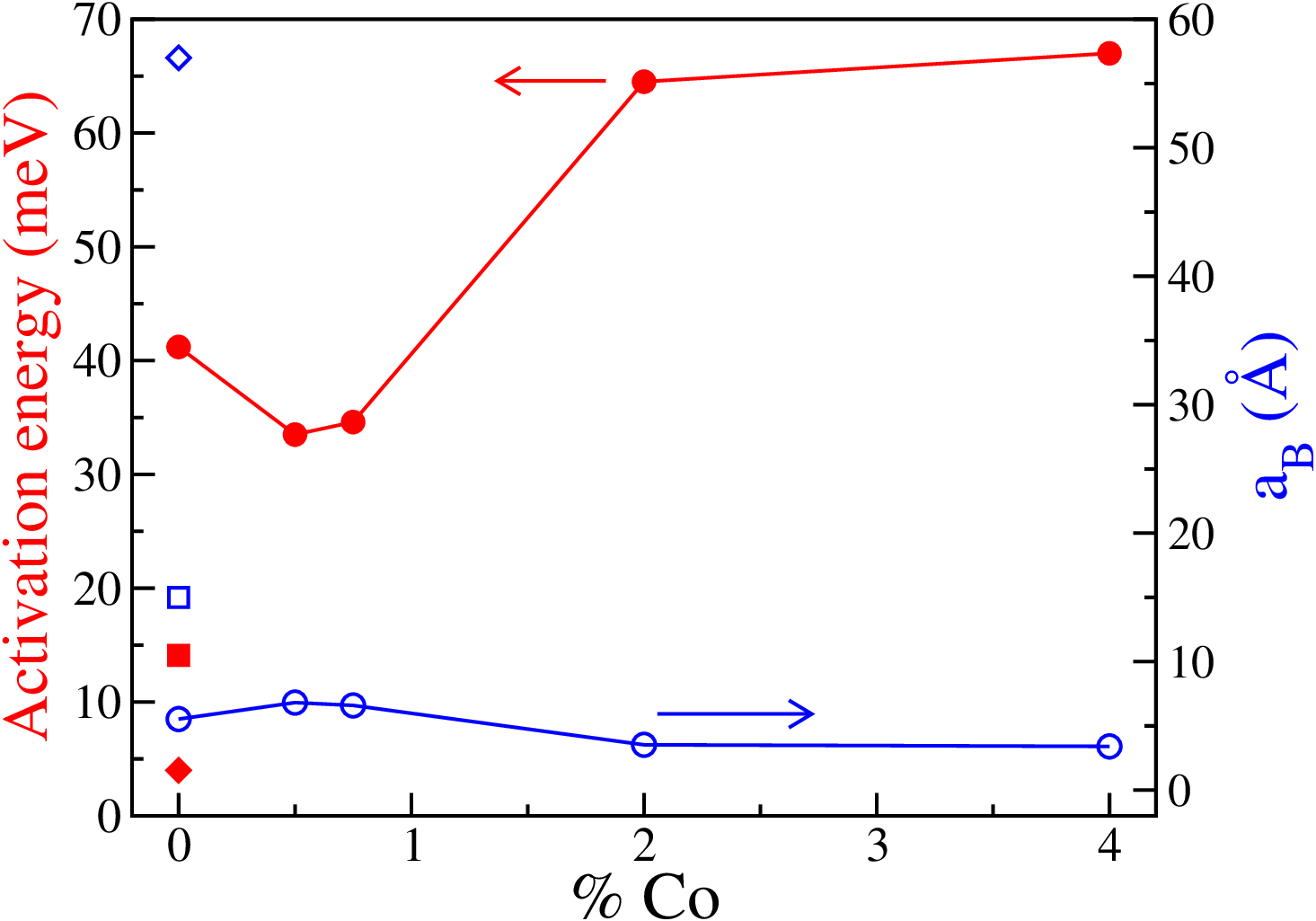}
        \end{center}
        \caption{Activation energies (closed symbols) and estimated localization radius
        $a_{\rm B}$ (open symbols) for undoped and Co-doped TiO$_2$. For these estimations,
        the static dielectric constant $\epsilon=31$ has been
        used. Circles are results from Ref.~\cite{shinde-un}, squares from Ref.~\cite{tang94}, and diamonds from Ref.~\cite{forro94}. }
\label{fig:activation}
\end{figure}

We argue here that, at high enough temperatures
compared to the carrier binding energy, carriers are thermally excited to conduction (or valence) bands from the impurity band becoming itinerant and mediating ferromagnetism by an effective
RKKY mechanism.
The carrier binding energy $\Delta E$ can be estimated by fitting the
resistivity curves to $\rho= \rho_0 \exp(\Delta E/k_{\rm B} T)$. The result
is unfortunately sample and composition dependent. For the undoped  TiO$_2$,
estimations go from $\sim 4$ meV~\cite{forro94} to $\sim 41$ 
meV~\cite{shinde-un}. In Ref.~\cite{tang94}, $a_{\rm B}$
in TiO$_2$
is estimated to be $15\, {\rm \AA}$ from the observation of an insulator to
metal transition upon doping at $n_c \sim 5\times 10^{18}$
cm$^{-3}$. This corresponds to $\Delta E= 14$ meV using the static
dielectric constant $\epsilon=31$~\cite{tang94}
and $m^*=m$. We show these results in Fig.~\ref{fig:activation}, together with the activation energy values corresponding to Co-doped TiO$_2$ films~\cite{shinde-un}. 
When doped with magnetic
ions, the resistivity increases by orders of magnitude proportionally
to the density of impurities, its slope is dramatically
enhanced, and does not show the insulator
to metal transition~\cite{ogale03,shinde03,toyosaki04} observed in the
undoped samples. The overall increase of the resistivity is due to the
scattering from charged impurities, as they produce strong disorder
in the system. However, the resistivity due to the impurity scattering is not
temperature dependent on the exponential scale of carrier activation. Therefore, $\Delta E$ can be calculated the same way as for undoped samples. 
The results for a particular series of Co-doped TiO$_2$~\cite{shinde-un} show that the activation energy increases upon doping (see Fig.~\ref{fig:activation}).  We expect these values to vary widely from sample to sample, as observed for the undoped compound. (This is also consistent with the current experimental situation where the observed $T_C$ in various nominally similar Co-TiO$_2$ samples varies widely from sample to sample.)

As the density of free carriers depends exponentially
on the impurity energy level (or the activation energy), a significant part of the
bound impurity band electrons can be promoted above the mobility edge at temperatures 
lower than $T_C$. This is particularly true in doped magnetic oxides,
where the claimed $T_C$ ($\approx 400-1000$K) is so high  that
the $k_{\rm B} T_C
>> \Delta E$ regime is easily reached producing a high density of
thermally activated mobile band carriers at   $k_{\rm B} T
\gtrsim \Delta E$.

\begin{figure}
\begin{center}
  \includegraphics*[width=3.0in]{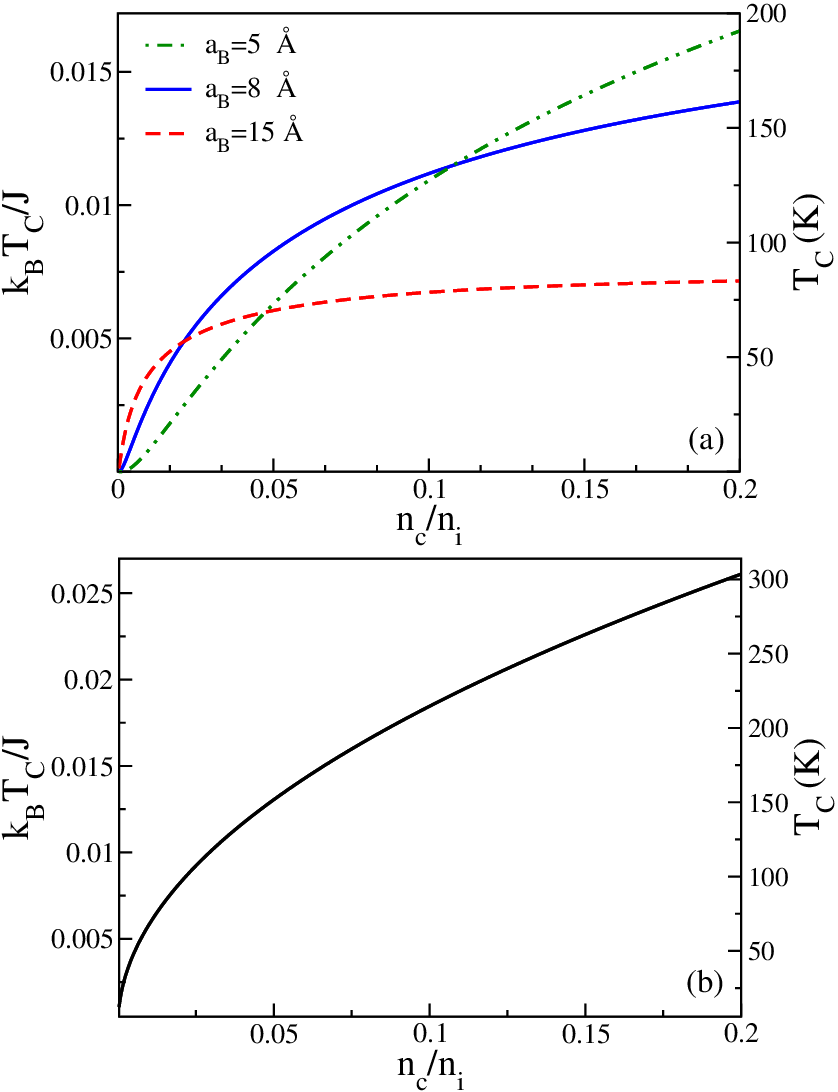}
  \end{center}
        \caption{(a) Estimation of ferromagnetic
       $T_C$ for the theory of bound magnetic polarons
       Eq.~(\ref{eq:Tc-perc}) and (b) for the RKKY model for itinerant
       carriers Eq.~(\ref{eq:Tc-nondeg}). $T_C$ on the right y-scale has been calculated using $J=1$
         eV. $S=1$, $a_o=3.23$ \AA~\cite{simpson04}, and
       $n_i=3 \,\times \, 10^{21} $ cm$^{-3}$ 
       (equivalent to $x=0.1$ for doped TiO$_2$).
	}
\label{fig:temp-nc}
\end{figure}

The estimation of the critical temperature depends on many variables:
(i) the density of carriers $n_c$, (ii) the density of magnetic ions $n_i$, (iii) the
exchange
coupling $J$ and, (iv)
in the bound magnetic polaron theory, the localization radius $a_{\rm B}$.
These parameters are either not precisely known or present a huge sample-to-sample variation:
(i) The
density of carriers $n_c$ has been measured by Hall effect giving results
ranging 
from $10^{18}$ to $10^{22}$ cm$^{-3}$~\cite{higgins-HE-04}, 
as the density of oxygen vacancies varies greatly with
the growth conditions. 
(ii) The absolute density of magnetic ions is known ($x=0.1$
corresponds to $n_i \sim 3 \times 10^{21}$ cm$^{-3}$ in TiO$_2$) but we
cannot estimate how many of those are magnetically active and,
therefore, relevant to long-range carrier-mediated
ferromagnetism. In fact, the magnetic moment per magnetic ion has a
strong dependence on sample characteristics~\cite{coey05}. 
This could be due to different effective values of magnetically 
active impurities.
In particular, for Co-doped TiO$_2$, $S$ is usually close to the low spin state $S=1/2$.
(iii) $J$ is not known in general though there are some estimates
which place its value above $1$ eV for ZnO~\cite{coey05}.  (iv) $a_{\rm B}$ calculated from the activation energies are shown in
Fig.~\ref{fig:activation} (right axes). Note the strong variation of
$a_{\rm B}$ from sample to sample in the undoped case.  This dramatically affects the
estimates of the critical temperature within the polaron percolation
model as shown in Fig.~\ref{fig:temp-nc}, as $T_C^{\rm perc}$ depends exponentially on  $a_{\rm B}^3 n_c$.

Estimations of $T_C$ from Eqs.~(\ref{eq:Tc-perc}) and
(\ref{eq:Tc-nondeg}) depend strongly
on all these unknown parameters and, consequently, $T_C$ can vary from
tens to hundreds of Kelvin upon tuning their values.
This
is illustrated in Fig.~\ref{fig:temp-nc}.  The local moment is taken to
be $S=1$, however, the moment per magnetic ion is another quantity
that is very sample dependent. A value of
$n_i=3 \, \times \, 10^{21}$ cm$^{-3}$ has been used. $T_C^{\rm perc}$ 
[Eq.~(\ref{eq:Tc-perc})] increases with the magnetic impurity density as $\sqrt{n_i}$ while $T_C^{\rm RKKY}$
[Eq.~(\ref{eq:Tc-nondeg})] only depends on the product  $\sqrt{n_i n_c}$.
Note
that both estimates of $T_C$ are the same order of magnitude and
close to experimental data.
However, due to the strong dependence of $T_C$ on unknown parameters,
we should not use the calculated value of $T_C$ as the sole criterion to elucidate the applicability of a
particular model. Rather, we should look at other evidence given by
experiment such as trends in magnetization curves and
magnetoresistance. We note, however, that we cannot rule out the
possibility that the strong variation in $T_C$ between different
experimental groups (or even from sample to sample in the same
group) arises precisely from the variation of the sample parameters
$n_i$, $n_c$, $a_{\rm B}$, etc. which will indeed lead to large $T_C$ variation! Further experiments are clearly needed to settle this important issue.

The measured magnetization versus temperature curves $M(T)$ 
usually present a very constant signature within a wide range 
at low temperatures~\cite{matsumoto01,ogale03,shinde03,chambers01},
though in some cases, 
a concave shape is observed~\cite{rode03,wang05}. As explained in
Sec.~\ref{sec:model}, this concave shape is expected for
insulating systems within the polaron percolation approach and for low
carrier density systems within the RKKY approach.

Magnetoresistance is very sample and composition dependent and there
are not many reports in the literature.  
In general, it is only
significant at temperatures much lower than $T_{C}$~\cite{matsumoto01,ogale03,jin01}. For doped ZnO, MR shows
different 
signs for different dopants and temperatures~\cite{jin01}, though these
results may not be relevant as these samples were not ferromagnetic.
Co-SnO$_2$ displays positive MR that becomes negative with increasing
$T$~\cite{ogale03}. 
Anatase Co-TiO$_2$ has positive MR for $T \le 5$ K ($60\%$ at
$T=2$K and $H=8$T)~\cite{matsumoto01}, while rutile Co-TiO$_2$
shows negative MR  up to almost room temperature (maximum $\sim -0.4\%$ at $T=100$K and $H=8$T)~\cite{toyosaki04}. The latter also presents an increase of MR for the
least resistive samples, the ones with higher carrier density.

As explained above, the polaron percolation model can explain a
negative MR, whose value depends on the change of $\langle S_z
\rangle$ with the applied magnetic field. At low temperatures, $\langle S_z
\rangle$ should already be saturated, therefore, MR is bound to be
small, consistent with results in O-DMS. On the other hand, 
both the percolation and RKKY models imply a maximum MR at $T_C$ when spin
fluctuations and 
susceptibility are very large and a small magnetic field can affect
magnetization dramatically. However, this behavior has not been found
in O-DMS. 

\section{Discussion} 
\label{sec:discussion}
In this work we are trying to understand the physical origin of
ferromagnetism in magnetically doped oxide materials
(e.g. Co-TiO$_2$) by assuming the underlying magnetic mechanism to be
carrier-mediated, qualitatively similar to that operational in the
widely studied DMS systems such as Ga$_{1-x}$Mn$_x$As. We use an
effective model of a local exchange coupling between the localized
magnetic moments (i.e. Co) and the localized carriers (since the
system, e.g. Co-TiO$_2$, is an insulator at $T=0$). We first show that
the very high $T_C$ (well above the room temperature) reported in the
literature cannot be understood entirely within the BMP percolation
picture {\it by itself} without fine-tuning the effective parameters
(e.g. exchange coupling, carrier and magnetic moment densities) to an
unreasonable degree. We therefore disagree with the recent attempts~\cite{coey05} in the literature to attribute FM in O-DMS entirely to BMP percolation. In general, the BMP percolation theory, which is
very natural for the insulating magnetic oxides and has already been
invoked in the literature~\cite{venkatesan04,coey04,coey05}, leads to
low values of $T_C$ for any reasonable assumptions about the system
parameters. The observed convex shape of the magnetization curves also
argues against an entirely BMP percolation model for magnetic oxide
ferromagnetism. In addition,
for TiO$_2$, the high temperature resistivity measurements
indicate the existence of a substantial thermally activated
carrier population consistent with RKKY.

We therefore suggest a composite scenario by taking into account the
relatively small carrier binding energy $\Delta E$ in these
materials. We propose that the low temperature BMPs in the system (in
the ferromagnetic state) become mobile (instead of going into a
disordered paramagnetic state through the percolation transition), and
these mobile carriers then produce long-range ferromagnetic coupling
among the magnetic impurities through the standard RKKY-Zener
mechanism, which could explain the experimentally observed
high-temperature ferromagnetism. Wide variations in effective carrier
($n_c$) and magnetic moment ($n_i$) densities due to different growth
conditions could easily accomodate the observed large variation in
$T_C$ among different groups.

We emphasize that our model of combined polaron percolation (at low
temperatures, where the carriers are still bound) and RKKY-Zener (at
high temperatures, where the carriers are activated into mobile
 states) mechanism seems to be the only reasonable
possibility for the ferromagnetism with high $T_C$ of the doped magnetic oxides
assuming that the ferromagnetism {\it is both intrinsic and
  carrier-mediated}. This combined model applies
mainly to TiO$_2$ where high temperature resistivity is activated~\cite{shinde03,toyosaki04,higgins-HE-04}. On the other hand, recent measurements
on Co-doped ZnO seem to rule out RKKY in favor of an
exclusive BMP mechanism~\cite{kittilstved06}. There are obviously other possibilities for the origin of ferromagnetism, which
are less interesting, but {\it not} necessarily less plausible.
First, the oxide ferromagnetism (in Co-TiO$_2$, Co-ZnO, etc.) could be
entirely extrinsic (and, some of it most likely is), arising from
contaminants (as happened, for example,  in the ferromagnetism of
calcium hexaborides) or from magnetic
nanoclustering~\cite{shinde03}. The Co nanoclusters in Co-TiO$_2$,
for example, could produce long-range ferromagnetic order through
dipolar coupling.  A second
intrinsic possibility for oxide ferromagnetism, increasingly gaining
ground on the basis of recent first principles band structure
calculations, is that the ordering in insulating
samples or regions arises from the
standard superexchange interaction between the localized magnetic
impurities (e.g. Co in Co-TiO$_2$)~\cite{janisch06}.
These local interactions would still
coexist with long-range carrier-mediated ferromagnetism. Other
calculations~\cite{ye06} find a $p-d$ hopping mechanism, not related
to free carriers but to clustering of the magnetic ions.
These possibilities certainly cannot
be ruled out, particularly in ordered magnetic oxides where such band
structure supercell calculations would, in principle, be
applicable. But the first principles calculations completely neglect
disorder, and cannot really explain the wide variation in magnetic
properties (e.g. $T_C$) seen in different laboratories. In addition, none of these posibilities 
(superexchange, p-d hopping mechanisms, or extrinsic nanocluster ferromagnetism) are carrier-mediated in the sense
of DMS materials, and as such are out of scope for our interest.

What we have discussed in this work is that, within the assumption of an
intrinsic carrier-mediated ferromagnetic mechanism (related to the
mechanisms operational in DMS materials), the most likely origin for
ferromagnetism in DMS oxides is a combination of BMP percolation at
low temperatures and RKKY-Zener coupling through activated carriers at
high temperatures. Only detailed experimental work can establish the
origin of ferromagnetism in magnetically doped oxides --- theory can
only suggest interesting possibilities without ruling out alternative
mechanisms such as extrinsic ferromagnetism or direct
superexchange. Among the various experimental evidence supporting the
model of carrier-mediated ferromagnetism in magnetically doped oxides
are the observation of AHE~\cite{toyosaki04}, electric field induced modulation
of magnetization~\cite{zhao-FE-05}, and optical magnetic circular dichroism~\cite{ando01}. It is clear
that much more systematic experimental data showing the magnetic
properties as a function of carrier density (and carrier properties)
will be needed before the issue of definitively establishing the
origin of ferromagnetism in doped magnetic oxides can be settled.

\section{Conclusion}
\label{sec:conclusion}
Before concluding, we emphasize that the new idea in this paper is that high-temperature RKKY ferromagnetism may be mediated in a semiconductor doped with magnetic impurities by thermally excited carriers in an otherwise-empty itinerant (conduction or valence) semiconductor band, in contrast to the usual band-carrier mediated RKKY ferromagnetism often discussed~\cite{priour04,dassarma04} in the context of  ferromagnetic (Ga,Mn)As where the valence band holes are thought to mediate the ferromagnetic RKKY coupling between the localized Mn magnetic moments. We believe that this RKKY coupling mediated by thermally excited carriers may be playing a role in the high observed $T_C$ in Co-doped TiO$_2$ where the experimentally measured conductivity is always activated in the ferromagnetic phase indicating the presence of substantial thermally activated free carriers.  Obviously, the necessary condition for such a thermally activated RKKY ferromagnetism is that electrical conduction in the system must be activated (and therefore insulating) in nature in contrast to metallic temperature-independent conductivity. If the observed conductivity in the system is temperature-independent, then the novel thermally-excited RKKY mechanism proposed by us simply does not apply.  Our motivation for suggesting this rather unusual thermally-activated RKKY DMS ferromagnetism has been the reported existence of very high transition temperatures in Co-doped TiO$_2$   which simply cannot be explained quantitatively by the bound magnetic polaron percolation picture of Kaminski and Das Sarma~\cite{kaminski02,dassarma-mag-03,kaminski03} although it is certainly possible that some of the oxide ferromagnetism arises purely from the BMP mechanism as has recently been argued~\cite{kittilstved06}.  The thermally activated RKKY ferromagnetic mechanism, while being necessary for high $T_C$, cannot be sufficient since at low temperatures the thermally activated carriers freeze out leading to an exponential suppression of the thermally activated RKKY ferromagnetism.  We therefore propose that BMP ferromagnetism~\cite{kaminski02,dassarma-mag-03,kaminski03} mediated by strongly localized carriers take over at low temperatures, as already proposed by Coey and collaborators~\cite{coey05} supplementing and complementing the high-temperature RKKY mechanism.  The two mechanisms coexist in the same sample with the high values of $T_C$ being controlled by the thermally activated RKKY mechanism and ferromagnetism persisting to $T=0$ due to the bound magnetic polaron percolation mechanism.  We note that in a single sample the two mechanisms will give rise to a unique $T_C$   depending on all the details of the system since the two mechanisms coexist whereas there could be considerable sample to sample $T_C$  variation due to the coexistence of the two mechanisms.  In our picture the high-temperature thermally activated RKKY mechanism smoothly interpolates to the low-temperature BMP ferromagnetism with a single transition temperature.  This picture of the coexistence of two complementary ferromagnetic mechanisms in oxides is essentially forced on us by our consideration of the possible transition temperatures achievable within the BMP model, which are just too low to explain the observations in (at least) the Co-doped TiO$_2$. 

Very recently, we have shown~\cite{re-entrant} that the confluence of two competing FM mechanisms, namely the BMP percolation in the impurity band at 'low' temperatures and the activated RKKY interaction in the conduction band at 'high' temperatures, could lead to an intriguing and highly non-trivial re-entrant FM transition in O-DMS where lowering temperature at first leads to a non-FM phase which gives way to a lower temperature second FM phase. A direct observation of our predicted~\cite{re-entrant} re-entrant FM would go a long way in validating the dual FM mechanism model introduced in this article.

To summarize, we have analyzed different proposed models for carrier-mediated
ferromagnetism in dilute magnetic oxides such as TiO$_2$, ZnO, andSnO$_2$. Due to the insulating character of these compounds,
a model based on the formation of bound magnetic polarons is
proposed. However, the binding energy of the electrons on the oxygen
vacancies that act as shallow donors is not large enough to keep the
electrons bound up to the high temperature reported for the $T_C$ ($\sim 700$K). Therefore, we propose that, at sufficiently
high temperatures, still below $T_C$, thermally excited carriers
also mediate ferromagnetism via an RKKY mechanism, complementing the
bound polaron picture and allowing a considerable enhancement of $T_C$.

We thank S. Ogale for valuable discussions.
This work is supported by the NSF, US-ONR, and NRI-SWAN.

\end{document}